\documentstyle[12pt]{article}
\newlength{\extraspace}
\newlength{\extraspaces}
\newcommand{\be}{\begin{equation}
\addtolength{\abovedisplayskip}{\extraspaces}
\addtolength{\belowdisplayskip}{\extraspaces}
\addtolength{\abovedisplayshortskip}{\extraspace}
\addtolength{\belowdisplayshortskip}{\extraspace}}
\newcommand{\ee}{\end{equation}}
\newcommand{\ba}{\begin{eqnarray}
\addtolength{\abovedisplayskip}{\extraspaces}
\addtolength{\belowdisplayskip}{\extraspaces}
\addtolength{\abovedisplayshortskip}{\extraspace}
\addtolength{\belowdisplayshortskip}{\extraspace}}
\newcommand{\ea}{\end{eqnarray}}
\newcommand{\nonu}{\nonumber \\[.5mm]}
\newcommand{\A}{&\!\!\!}
\begin{document}
\thispagestyle{empty}
\begin{flushright}
SIT-LP-08/01 \\
January, 2008
\end{flushright}
\vspace{7mm}
%
%
\begin{center}
{\large{\bf On $N = 2$ superfield for $N = 2$ vector supermultiplet \\[2mm]
in two dimensional spacetime}} \\[20mm]
{\sc Kazunari Shima}
\footnote{
\tt e-mail: shima@sit.ac.jp} \ 
and \ 
{\sc Motomu Tsuda}
\footnote{
\tt e-mail: tsuda@sit.ac.jp} 
\\[5mm]
{\it Laboratory of Physics, 
Saitama Institute of Technology \\
Fukaya, Saitama 369-0293, Japan} \\[20mm]
\begin{abstract}
We focus on the superfield formulation for a $N = 2$ vector supermultiplet 
in two dimensional spacetime and explicitly show that the Wess-Zumino gauge condition 
for a $N = 2$ superfield is compatible with familiar SUSY (plus $U(1)$ gauge) transformations 
for the vector supermultiplet. 
$N = 2$ SUSY invariant mass and Yukawa interaction terms for the vector supermultiplet 
are also constructed from the superfield explicitly in addition to a free (kinetic) action. 
\\[5mm]
\end{abstract}
\end{center}

\newpage

\noindent
Superfield formulation in superspace gives systematic understandings 
for various supersymmetric (SUSY) field theories 
(for further references on superfields, see \cite{WB,PW}). 
In two dimensional spacetime ($d = 2$), superfields were constructed in the context 
of discussing classical solutions for SUSY field theories \cite{DVF}, 
e.g. $N = 1$ and $N = 2$ superfields for vector supermultiplets 
are given as a spinor and a scalar, respectively. 
Moreover, a realistic SUSY model can be constructed by $N \ge 2$ SUSY 
in the SGM scenario \cite{STL}, 
where we have found \cite{ST1,ST2} the relation (equivalence) 
between a nonlinear (NL) SUSY model \cite{VA} 
and linear SUSY interacting field theories in $d = 2$. 
In order to study systematics in the linearization of $N \ge 2$ NLSUSY 
(for SUSY interacting field theories), 
it is important to know the details of $N \ge 2$ superfield formulation. 

In this letter, we focus on the superfield formulation 
for the $N = 2$ vector supermultiplet in $d = 2$ and discuss on the basics in detail. 
Based on the $d = 2$, $N = 2$ superfield \cite{DVF}, we explicitly show that 
the Wess-Zumino (WZ) gauge condition is compatible with familiar SUSY 
(plus $U(1)$ gauge) transformations for the vector supermultiplet. 
$N = 2$ SUSY invariant mass and Yukawa interaction terms 
for the vector supermultiplet are also constructed from the superfield explicitly 
in addition to a free (kinetic) action. 

Let us first introduce the following $N = 2$ superfield in $d = 2$ \cite{DVF} 
on superspace coordinates $(x^a, \theta_\alpha^i)$, 
\footnote{
Minkowski spacetime indices in $d = 2$ are denoted by $a, b, \cdots = 0, 1$ 
and $SO(2)$ internal indices are $i, j, \cdots = 1, 2$. 
The Minkowski spacetime metric is 
${1 \over 2}\{ \gamma^a, \gamma^b \} = \eta^{ab} = {\rm diag}(+, -)$. 
As for the conventions in $d = 2$ for the $\gamma$ matrices etc., 
for example, see \cite{UZ}. 
}
\ba
{\cal V}(x, \theta^i) \A = \A C(x) + \bar\theta^i \Lambda^i(x) 
+ {1 \over 2} \bar\theta^i \theta^j M^{ij}(x) 
- {1 \over 2} \bar\theta^i \theta^i M^{jj}(x) 
+ {1 \over 4} \epsilon^{ij} \bar\theta^i \gamma_5 \theta^j \phi(x) 
\nonu
\A \A 
- {i \over 4} \epsilon^{ij} \bar\theta^i \gamma_a \theta^j v^a(x) 
- {1 \over 2} \bar\theta^i \theta^i \bar\theta^j \lambda^j(x) 
- {1 \over 8} \bar\theta^i \theta^i \bar\theta^j \theta^j D(x), 
\label{VSF}
\ea
where the component fields are denoted by $(C, D)$ for two scalar fields, 
$(\Lambda^i, \lambda^i)$ for four spinor fields, 
$\phi$ for a pseudo scalar field, $v^a$ for a vector field, 
and $M^{ij} = M^{(ij)}$ $\left(= {1 \over 2}(M^{ij} + M^{ji}) \right)$ 
for three scalar fields ($M^{ii} = \delta^{ij} M^{ij}$), 
respectively. 
Under superspace translations, 
\be 
x'^a = x^a + i \bar\zeta^i \gamma^a \theta^i, 
\ \ \ \theta'{}_\alpha^i = \theta_\alpha^i + \zeta_\alpha^i, 
\ee
the superfield (\ref{VSF}) transforms as 
\be
\delta_\zeta {\cal V}(x, \theta^i) = \bar\zeta^i Q^i {\cal V}(x, \theta^i) 
\label{VSFtransfn}
\ee
with supercharges 
\be
Q_\alpha^i = {\partial \over \partial\bar\theta^{\alpha i}} 
+ i \!\!\not\!\partial \theta_\alpha^i, 
\ee
satisfying $\{ Q_\alpha^i, Q_\beta^j \} = - 2 \delta^{ij} (\gamma^a C)_{\alpha \beta} P_a$. 
The superfield transformation (\ref{VSFtransfn}) 
gives SUSY transformations for the component fields as, 
\ba
\A \A 
\delta_\zeta C = \bar\zeta^i \Lambda^i, 
\nonu
\A \A 
\delta_\zeta \Lambda^i 
= - i \!\!\not\!\partial C \zeta^i + M^{ij} \zeta^j - M^{jj} \zeta^i 
+ {1 \over 2} \epsilon^{ij} \phi \gamma_5 \zeta^j 
- {i \over 2} \epsilon^{ij} v \cdot \gamma \zeta^j, 
\nonu
\A \A 
\delta_\zeta M^{12} = \bar\zeta^{(1} \lambda^{2)} 
- i \bar\zeta^{(1} \!\!\not\!\partial \Lambda^{2)}, 
\nonu
\A \A 
\delta_\zeta M^{11} = \bar\zeta^1 \lambda^1 + i \bar\zeta^2 \!\!\not\!\partial \Lambda^2, 
\ \ \ \delta_\zeta M^{22} = \bar\zeta^2 \lambda^2 + i \bar\zeta^1 \!\!\not\!\partial \Lambda^1, 
\nonu
\A \A 
\delta_\zeta \phi = \epsilon^{ij} (- \bar\zeta^i \gamma_5 \lambda^j 
- i \bar\zeta^i \gamma_5 \!\!\not\!\partial \Lambda^j), 
\nonu
\A \A 
\delta_\zeta v^a = \epsilon^{ij} (- i \bar\zeta^i \gamma^a \lambda^j 
- \bar\zeta^i \!\!\not\!\partial \gamma^a \Lambda^j), 
\nonu
\A \A 
\delta_\zeta \lambda^i = - i \!\!\not\!\partial M^{ij} \zeta^j 
- {i \over 2} \epsilon^{ij} \gamma_5 \!\!\not\!\partial \phi \zeta^j 
- {1 \over 2} \epsilon^{ij} \gamma_a \!\!\not\!\partial v^a \zeta^j 
+ D \zeta^i, 
\nonu
\A \A 
\delta_\zeta D = - i \bar\zeta^i \!\!\not\!\partial \lambda^i. 
\label{SUSY}
\ea

Next following the ref.\cite{WB} we define the SUSY generalization 
of a gauge transformation of ${\cal V}(x,\theta^i)$ as 
\be
\delta_g {\cal V} = \Phi^1 + \alpha \Phi^2, 
\label{gauge}
\ee
where $\alpha$ is an arbitrary real parameter, 
and $\Phi^i$ are generalized gauge parameters in the form of $N = 2$ scalar superfields 
for a $N = 2$ scalar supermultiplet \cite{UZ}, 
\ba
\Phi^i(x, \theta^i) \A = \A B^i(x) + \bar\theta^i \chi(x) - \epsilon^{ij} \bar\theta^j \nu(x) 
- {1 \over 2} \bar\theta^j \theta^j F^i(x) + \bar\theta^i \theta^j F^j(x) 
- i \bar\theta^i \!\!\not\!\partial B^j(x) \theta^j 
\nonu
\A \A 
+ {i \over 2} \bar\theta^j \theta^j (\bar\theta^i \!\!\not\!\partial \chi(x) 
- \epsilon^{ik} \bar\theta^k \!\!\not\!\partial \nu(x)) 
+ {1 \over 8} \bar\theta^j \theta^j \bar\theta^k \theta^k \Box B^i(x) 
\ea
with the component fields being denoted by $B^i$ for two scalar fields, 
$(\chi, \nu)$ for two spinor fields and $F^i$ for two auxiliary scalar fields. 
(The gauge transformation (\ref{gauge}) may be recasted by using the ordinary 
(covariant) derivative as for the spinor-superfield case \cite{DVF}.) 
The explicit component form of the gauge transformation (\ref{gauge}) is 
\ba
\A \A 
\delta_g C = B^1 + \alpha B^2, 
\nonu
\A \A 
\delta_g \Lambda^1 = \chi + \alpha \nu, 
\ \ \ \delta_g \Lambda^2 = \alpha \chi - \nu, 
\nonu
\A \A 
\delta_g M^{12} = \alpha F^1 + F^2, 
\nonu
\A \A 
\delta_g M^{11} = F^1 - \alpha F^2, 
\ \ \ \delta_g M^{22} = - F^1 + \alpha F^2, 
\nonu
\A \A 
\delta_g \phi = 0, 
\nonu
\A \A 
\delta_g v^a = - 2 \partial^a (\alpha B^1 - B^2), 
\nonu
\A \A 
\delta_g \lambda^1 = - i \!\!\not\!\partial (\chi + \alpha \nu), 
\ \ \ \delta_g \lambda^2 = - i \!\!\not\!\partial (\alpha \chi - \nu), 
\nonu
\A \A 
\delta_g D = - \Box (B^1 + \alpha B^2). 
\label{gauge-comp}
\ea
Note that $v^a$ transforms as an Abelian gauge field 
and, in addition to $\phi$, $M^{ii}$ is a gauge invariant quantity ($\delta_g M^{ii} = 0$). 
By using the gauge freedoms of ($B^i$, $\chi$, $\nu$, $F^i$) in Eq.(\ref{gauge-comp}), 
we take the WZ gauge \cite{DVF}, 
\be
C = 0, \ \ \ \Lambda^i = 0, \ \ \ M^{12} = 0, \ \ \ M^{11} - M^{22} = 0, 
\label{WZ}
\ee
and the superfield (\ref{VSF}) is rewritten in terms of $A = M^{ii}$. 
Then, the generalized gauge transformations (\ref{gauge-comp}) 
become the ordinary ones, 
\ba
\A \A 
\delta_g A = \delta_g M^{ii} = 0, 
\nonu
\A \A 
\delta_g \phi = 0, 
\nonu
\A \A 
\delta_g v^a = - 2 \partial^a (\alpha B^1 - B^2), 
\nonu
\A \A 
\delta_g \lambda^i = 0, 
\nonu
\A \A 
\delta_g D = 0, 
\ea
where the gauge parameters of $v^a$ are essentially one parameter 
because one of them is used by $C = 0$ in Eq.(\ref{WZ}). 
The superfield (\ref{VSF}) also reduces to \cite{NO}
\ba
V(x, \theta^i) \A = \A {\cal V}(x, \theta^i) \vert_{\rm WZ\ gauge} 
\nonu
\A = \A - {1 \over 4} \bar\theta^i \theta^i A(x) 
+ {1 \over 4} \epsilon^{ij} \bar\theta^i \gamma_5 \theta^j \phi(x) 
- {i \over 4} \epsilon^{ij} \bar\theta^i \gamma_a \theta^j v^a(x) 
- {1 \over 2} \bar\theta^i \theta^i \bar\theta^j \lambda^j(x) 
\nonu
\A \A 
- {1 \over 8} \bar\theta^i \theta^i \bar\theta^j \theta^j D(x). 
\label{VSF-WZ}
\ea

The WZ gauge (\ref{WZ}) is violated by the SUSY transformations (\ref{SUSY}). 
Therefore, we consider modified SUSY transformations (for example, see \cite{MSS}), 
\be
\tilde \delta_\zeta = \delta_\zeta + \delta_g. 
\label{mSUSY}
\ee
in order to maintain the WZ gauge condition under Eq.(\ref{mSUSY}). 
For the eliminated component fields by Eq.(\ref{WZ}), 
the modified SUSY transformations (\ref{mSUSY}) are written as 
\ba
\A \A 
\tilde \delta_\zeta C = B^1 + \alpha B^2, 
\nonu
\A \A 
\tilde \delta_\zeta \Lambda^1 
= - A \zeta^1 + M^{11} \zeta^1 
+ {1 \over 2} \phi \gamma_5 \zeta^2 
- {i \over 2} v \cdot \gamma \zeta^2 
+ \chi + \alpha \nu, 
\nonu
\A \A 
\tilde \delta_\zeta \Lambda^2 
= - A \zeta^2 + M^{22} \zeta^2 
- {1 \over 2} \phi \gamma_5 \zeta^1 
+ {i \over 2} v \cdot \gamma \zeta^1 
+ \alpha \chi - \nu, 
\nonu
\A \A 
\tilde \delta_\zeta M^{12} 
= \bar\zeta^{(1} \lambda^{2)} + \alpha F^1 + F^2. 
\nonu
\A \A 
\tilde \delta_\zeta (M^{11} - M^{22}) 
= \bar\zeta^1 \lambda^1 - \bar\zeta^2 \lambda^2 
+ 2 (F^1 - \alpha F^2), 
\ea
and we impose 
\be
\tilde \delta_\zeta C = 0, \ \ \ \tilde \delta_\zeta \Lambda^i = 0, 
\ \ \ \tilde \delta_\zeta M^{12} = 0, 
\ \ \ \tilde \delta_\zeta (M^{11} - M^{22}) = 0, 
\label{WZSUSY}
\ee
i.e. 
\ba
\A \A 
B^1 + \alpha B^2 = 0, 
\nonu
\A \A 
\chi + \alpha \nu = A \zeta^1 - M^{11} \zeta^1 
- {1 \over 2} \phi \gamma_5 \zeta^2 + {i \over 2} v \cdot \gamma \zeta^2, 
\nonu
\A \A 
\alpha \chi - \nu = A \zeta^2 - M^{22} \zeta^2 
+ {1 \over 2} \phi \gamma_5 \zeta^1 
- {i \over 2} v \cdot \gamma \zeta^1, 
\nonu
\A \A 
\alpha F^1 + F^2 = - \bar\zeta^{(1} \lambda^{2)}, 
\nonu
\A \A 
F^1 - \alpha F^2 = - {1 \over 2} (\bar\zeta^1 \lambda^1 - \bar\zeta^2 \lambda^2). 
\label{WZSUSY-comp}
\ea
Substituting these (in particular, the first three equations in Eq.(\ref{WZSUSY-comp})) 
into Eq.(\ref{mSUSY}) for the remaining component fields ($v^a$, $\lambda^i$, $A$, $\phi$, $D$) 
gives the familiar SUSY transformations for the $d = 2$, $N = 2$ vector supermultiplet \cite{ST1} 
plus the $U(1)$ gauge transformation of $v^a$, 
\ba
\A \A 
\tilde \delta_\zeta A \ (= \tilde \delta_\zeta M^{ii}) = \bar\zeta^i \lambda^i, 
\nonu
\A \A 
\tilde \delta_\zeta \phi = - \epsilon^{ij} \bar\zeta^i \gamma_5 \lambda^j, 
\nonu
\A \A 
\tilde \delta_\zeta v^a = - i \epsilon^{ij} \bar\zeta^i \gamma^a \lambda^j 
- 2 \partial^a (\alpha B^1 - B^2), 
\nonu
\A \A 
\tilde \delta_\zeta \lambda^i = (D - i \!\!\not\!\partial A) \zeta^i 
+ {1 \over 2} \epsilon^{ab} \epsilon^{ij} F_{ab} \gamma_5 \zeta^j 
- i \epsilon^{ij} \gamma_5 \!\!\not\!\partial \phi \zeta^j, 
\nonu
\A \A 
\tilde \delta_\zeta D = - i \bar\zeta^i \!\!\not\!\partial \lambda^i, 
\label{mSUSY-comp}
\ea
where $F_{ab} = \partial_a v_b - \partial_b v_a$, 
and $-2(\alpha B^1 - B^2)$ is the gauge parameter for the $U(1)$ gauge. 
The SUSY transformations (\ref{mSUSY-comp}) satisfy a closed commutator algebra 
\be
[ \tilde \delta_{\zeta_1}, \tilde \delta_{\zeta_2} ] = \delta_P(\Xi^a) + \delta_g(\theta), 
\label{N2D2commg}
\ee
where $\delta_P(\Xi^a)$ means a translation with a generator 
$\Xi^a = 2 i \bar\zeta_1^i \gamma^a \zeta_2^i$, 
and $\delta_g(\theta)$ is the $U(1)$ gauge transformation
with a generator $\theta = - 2 (i \bar\zeta_1^i \gamma^a \zeta_2^i$ $v_a 
- \epsilon^{ij} \bar\zeta_1^i \zeta_2^j A 
- \bar\zeta_1^i \gamma_5 \zeta_2^i \phi)$, which appears in the commutator for $v^a$. 
As is shown below, actions (Eqs. from (\ref{V0}) to (\ref{Vf})) constructed 
from the superfield (\ref{VSF-WZ}) are also SUSY and $U(1)$-gauge invariant under Eq.(\ref{mSUSY-comp}). 
Therefore, the superfield (\ref{VSF}) reduces to Eq.(\ref{VSF-WZ}) in the WZ gauge (\ref{WZ}) 
with the well-defined SUSY and $U(1)$ gauge transformations (\ref{mSUSY-comp}) 
without violating the valance between the bosonic and fermionic degrees of freedom. 
In the above arguments, when $\alpha = 1$ the symmetrization of the indices for $M^{12}$ 
is manifest, while when $\alpha = \pm i$ the $U(1)$ gauge transformation 
is not induced in Eq.(\ref{mSUSY-comp}). 

Let us evaluate SUSY invariant (renormalizable) actions 
for the $N = 2$ vector supermultiplet \cite{ST1}, 
which are constructed from the superfield (\ref{VSF-WZ}) 
with the component fields $(v^a, \lambda^i, A, \phi, D)$. 
By using differential operators in $N = 2$ superspace, 
\be
D_\alpha^i = {\partial \over \partial\bar\theta^{\alpha i}} 
- i \!\!\not\!\partial \theta_\alpha^i, 
\ee
we define the following scalar and pseudo scalar superfields, 
\be
W^{ij} = \bar D^i D^j V, \ \ \ W_5^{ij} = \bar D^i \gamma_5 D^j V. 
\ee
Then, the free (kinetic) action is obtained from $W^{ij}$ and $W_5^{ij}$ as 
\ba
S_{V 0} \A = \A {1 \over 4} \int d^2 x 
\left[ 
\int d^2 \theta^i \ {1 \over 8} (\overline{D^j W^{kl}} D^j W^{kl} 
+ \overline{D^j W_5^{kl}} D^j W_5^{kl}) 
+ {1 \over 4} \int d^4 \theta^i \ {8 \over \kappa} \xi V 
\right]_{\theta^i = 0} 
\nonu
\A = \A \int d^2 x \left\{ - {1 \over 4} (F_{ab})^2 
+ {i \over 2} \bar\lambda^i \!\!\not\!\partial \lambda^i 
+ {1 \over 2} (\partial_a A)^2 + {1 \over 2} (\partial_a \phi)^2 
+ {1 \over 2} D^2 - {1 \over \kappa} \xi D 
\right\}, 
\label{V0}
\ea
where the last term means the Fayet-Iliopoulos $D$-term 
with an arbitrary dimensionless paramater $\xi$. 
On the other hand, the quadratic and cubic terms of $W^{ij}$ and $W_5^{ij}$ 
leads to the mass and Yukawa interaction terms as follows: 
\ba
S_{Vm} \A = \A {1 \over 4} \int d^2 x \ m \left[ 
\int d^2 \theta^i \{ (W^{jk})^2 + (W_5^{jk})^2 \} 
+ \int d \bar\theta^i d \theta^j 
(W^{ik} W^{jk} + W_5^{ik} W_5^{jk}) 
\right]_{\theta^i = 0} 
\nonu
\A = \A \int d^2 x 
\left\{ - {1 \over 2} m \ 
( \bar\lambda^i \lambda^i - 2 AD + \epsilon^{ab} \phi F_{ab} ) \right\}, 
\label{Vm}
\\
S_{Vf} \A = \A {1 \over 4} \int d^2 x \ f \left[ 
{1 \over 2} \int d^2 \theta^i W^{jk} (W^{jl} W^{kl} + W_5^{jl} W_5^{kl}) 
\right. 
\nonu
\A \A 
\left. 
+ \int d \bar\theta^i d \theta^j \{ W^{ij} (W^{kl} W^{kl} + W_5^{kl} W_5^{kl}) 
+ W^{ik} (W^{jl} W^{kl} + W_5^{jl} W_5^{kl}) \} 
\right]_{\theta^i = 0} 
\nonu
\A = \A \int d^2 x 
\{ f ( A \bar\lambda^i \lambda^i + \epsilon^{ij} \phi \bar\lambda^i \gamma_5 \lambda^j 
- A^2 D + \phi^2 D + \epsilon^{ab} A \phi F_{ab} ) \}, 
\label{Vf}
\ea
where $f$ is an arbitrary constant with the dimension (mass)$^1$. 
The actions (\ref{V0}), (\ref{Vm}) and (\ref{Vf}) are invariant 
under the $N = 2$ SUSY and $U(1)$ gauge transformations (\ref{mSUSY-comp}), respectively. 

To summarize, we have shown that the superfield (\ref{VSF}) consistently reduces to 
Eq.(\ref{VSF-WZ}) with the minimal (off-shell) set of the comonent fields 
$(v^a, \lambda^i, A, \phi, D)$ in the WZ gauge (\ref{WZ}). 
The condition (\ref{WZSUSY}), i.e. Eq.(\ref{WZSUSY-comp}) 
to maintain the WZ gauge (\ref{WZ}) under the modified SUSY transformations (\ref{mSUSY}) 
gives the familiar $N = 2$ SUSY and $U(1)$ gauge transformations (\ref{mSUSY-comp}). 
The free (kinetic) action, the mass and Yukawa interaction terms 
for the $N = 2$ vector supermultiplet have been obtained 
from the superfield (\ref{VSF-WZ}) as in Eqs. from (\ref{V0}) to (\ref{Vf}). 
The extension of this $N = 2$ superfield formulation to $d = 4$ 
is an important problem which is now in progress. 
Also the coupling of matter supermultiplets to the $N = 2$ superfield (\ref{VSF}) 
is an interesting problem under study.

\newpage

%
\newcommand{\NP}[1]{{\it Nucl.\ Phys.\ }{\bf #1}}
\newcommand{\PL}[1]{{\it Phys.\ Lett.\ }{\bf #1}}
\newcommand{\CMP}[1]{{\it Commun.\ Math.\ Phys.\ }{\bf #1}}
\newcommand{\MPL}[1]{{\it Mod.\ Phys.\ Lett.\ }{\bf #1}}
\newcommand{\IJMP}[1]{{\it Int.\ J. Mod.\ Phys.\ }{\bf #1}}
\newcommand{\PR}[1]{{\it Phys.\ Rev.\ }{\bf #1}}
\newcommand{\PRL}[1]{{\it Phys.\ Rev.\ Lett.\ }{\bf #1}}
\newcommand{\PTP}[1]{{\it Prog.\ Theor.\ Phys.\ }{\bf #1}}
\newcommand{\PTPS}[1]{{\it Prog.\ Theor.\ Phys.\ Suppl.\ }{\bf #1}}
\newcommand{\AP}[1]{{\it Ann.\ Phys.\ }{\bf #1}}

\end{document}